\documentclass[12pt]{iopart}
\usepackage{iopams} 
\usepackage{graphicx}
\usepackage{multirow}
\begin{document}
\title{One-electron atomic-molecular ions containing Lithium in a strong magnetic field}
\author{H Olivares Pil\'on$^1$$^{,*}$, D Baye$^2$$^{,\dag}$, A V Turbiner$^1$$^{,\ddag}$ and J C L\'opez Vieyra$^1$$^{,\S}$}
\address{$^1$ Instituto de Ciencias Nucleares, UNAM, Apartado Postal 70-543, M\'exico DF, 04510 M\'exico }
\address{$^2$ Physique Quantique, CP 165/82, Universit\'e Libre de Bruxelles, B 1050 Brussels, Belgium}
\eads{\mailto{$^{*}$horop@nucleares.unam.mx}, $^{\dag}$\mailto{dbaye@ulb.ac.be}, $^{\ddag}$\mailto{turbiner@nucleares.unam.mx}, $^{\S}$\mailto{vieyra@nucleares.unam.mx} }
\date{today}
\begin{abstract}
The one-electron Li-containing Coulomb systems of atomic type $(li, e)$ and molecular type $(li, li, e)$,  $(li, \alpha, e)$ and $(li, p, e)$  are studied in the presence of a strong magnetic field $B \leq 10^{7}$ a.u.\ in the non-relativistic framework.
They are considered at the Born-Oppenheimer approximation of zero order (infinitely massive centers) within the parallel configuration (molecular axis parallel to the magnetic field). The variational and Lagrange-mesh methods are employed in complement to each other. It is demonstrated that the molecular systems ${\rm LiH}^{3+}$, ${\rm LiHe}^{4+}$ and ${\rm Li}_{2}^{5+}$ can exist for sufficiently strong magnetic fields $B \gtrsim  10^{4}$ a.u.\ and that ${\rm Li}_{2}^{5+}$ can even be stable at magnetic fields typical of magnetars.
\end{abstract}

\maketitle

\section{Introduction}

The existence of very strong magnetic fields in the atmosphere of neutron stars $B\sim 10^{12}\,$G ($\sim 10^3\,$a.u.)~\footnote{$1$ a.u. $ \approx 2.35\times10^{9}$~G$ = 2.35\times10^5$~T.} and magnetars $B\sim 10^{16}\,$G ($\sim 10^7\,$a.u.) has posed different questions about the structure of matter under the influence of such extreme magnetic fields (see for a review e.g. \cite{Lai:2001}). In particular, it raised the question about the possible atomic and molecular systems which can exist in a strong magnetic field: can molecular chains be formed? The interest in getting answers to these questions dramatically increased after the long-expected discoveries made by Chandra and XMM-Newton $X$-ray observatories of absorption features in the soft X-ray spectra of several magnetic stars. The first of that was the celebrated 2002 findings in the radiation coming from the isolated neutron star 1E1207.4-5209 of two wide absorption features at $\sim $0.7\ KeV and $\sim$1.4\ KeV \cite{chandra}. One of possible explanations of these features might be the presence of the traditional ${\rm H}_{2}^{+}$ and the exotic ${\rm H}_{3}^{2+}$ \cite{Turbiner:1999} molecular ions in the atmosphere of this neutron star \cite{Turbiner:2004NS}. It is one of the reasons to assume that other traditional and exotic systems - atomic and/or molecular, containing one or more electrons - might be present in the atmosphere of the neutron stars and magnetars. It has recently been predicted that a variety of hydrogenic, helium and mixed hydrogen/helium molecular ions of one and two electrons can be formed when a sufficiently strong magnetic field is imposed \cite{phrep,chains2009}.

It is already known that the simplest molecular ion ${\rm H}_{2}^{+}$ exists for all magnetic fields, while two more one-electron exotic ions ${\rm He}_2^{3+}$ and ${\rm HeH}^{2+}$ might begin to exist for magnetic fields $B\approx 10^2\,$a.u.\ and $B\approx 10^4\,$a.u., respectively (see \cite{Turbiner:2007}). Furthermore, for magnetic fields $B\gtrsim 10^4\,$ a.u.\ the exotic compound ${\rm He}_2^{3+}$ becomes the most bound one-electron system among the one-electron hydrogenic-helium ions. Then, a natural question arises: Can other one-electron two-center systems $(Z_1,Z_2,e)$ exist in a strong magnetic field?  A simple electrostatic analysis indicates that this is indeed possible for $Z_{1,2} < 4$ (see \cite{phrep}).

The goal of the present work is to study the existence of Lithium-containing ions  ${\rm Li}^{2+}$, ${\rm LiH}^{3+}$, ${\rm LiHe}^{4+}$ and ${\rm Li}_{2}^{5+}$ in the presence of a strong magnetic field. The relativistic corrections are assumed to be of a small importance for $B\le 10^{7}$\ a.u.\ following the analysis by Duncan \cite{Duncan} that the longitudinal electronic motion is still deeply non-relativistic. The molecular axis is assumed to be aligned parallel to the magnetic field line since it is rather evident that this configuration is optimal if a magnetic field is sufficiently strong \cite{phrep, BJS09}. The positively-charged centers are considered to be infinitely heavy (zero-order Born-Oppenheimer approximation).

We use two methods. The {\it variational method} with physically relevant trial functions and the {\it Lagrange-mesh method}.  The first one has been proven to be very efficient \cite{phrep} with physically motivated trial functions. On the other hand, the {\it Lagrange-mesh method} \cite{h2malla} nowadays provides the most accurate results for the total energy of the ${\rm H}_{2}^{+}$ ion.  However, in practice, the implementation of the Lagrange-mesh method is made easier with an {\it a priori} knowledge of the equilibrium distance and the corresponding total energy. Our approach is the following: as a first step we use the {\it variational method} to obtain  preliminary but already sufficiently accurate results for the equilibrium distance and the total energy and then the {\it Lagrange-mesh method} is used to check and possibly improve the variational results.

Atomic units $m_{e}= e = \hbar = 1$ are used throughout, although the energy is given in Rydbergs.

\section{One-electron molecular ion at the Born-Oppenheimer approximation}

\begin{figure}[!htp]
\centering
\includegraphics[height=5.0cm]{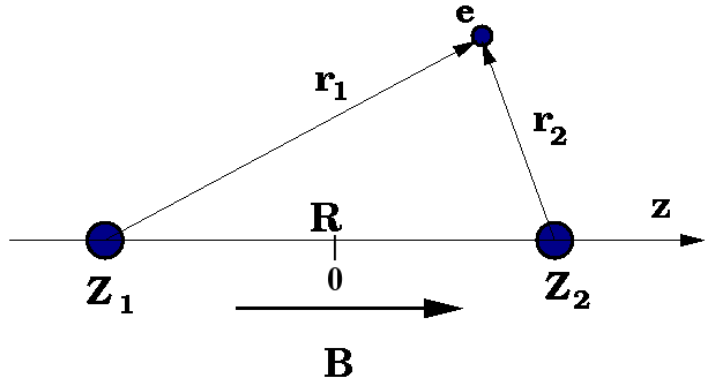}
\caption{Geometrical setting for the system of one-electron and two charged centers $Z_{1}$ and $Z_{2}$ placed in a magnetic field $B$  in parallel configuration along the $z$-axis. $R$ is the internuclear distance and $r_{1}$, $r_{2}$ are the distances between the electron and the charged centers $Z_{1}$ and $Z_{2}$, respectively.}
\label{fig:lix}
\end{figure}

The Hamiltonian which describes two infinitely heavy centers   of charges $Z_{1}$ and $Z_{2}$ situated along the $z$ axis, and one electron placed in a uniform constant magnetic field directed along the $z-$axis, $\nobreak{{\bf B}=(0,0,B)}$ is given by
\begin{equation}
 {\hat H} = -\Delta  -
\frac{2\,Z_1}{r_1} -\frac{2 Z_2}{r_2}\,+\frac{2\,Z_1 Z_2}{R}
+ ({\hat p} {\cal A}+{\cal A}{\hat p}) +  {\cal A}^2 \ ,
\end{equation}
(for geometrical setting see figure~\ref{fig:lix}, the origin of the coordinate system is the middle point between the nuclei for all molecular ions considered here). The vector potential corresponding to a constant magnetic field ${\bf  B}=(0,0,B)$ is chosen in the symmetric gauge
\begin{equation}
 {\cal A}= \frac{B}{2}(-y, \,x,\, 0)\ .
  \label{Vec}
\end{equation}
Then the Hamiltonian takes the form
\begin{equation}
 {\hat H} = -\Delta  -
\frac{2\,Z_1}{r_1} -\frac{2 Z_2}{r_2}\,+\frac{2\,Z_1 Z_2}{R}
+ B {\hat L}_{z} + \frac{B^{2}}{4}\rho^{2} \,,
\label{ham-LiX}
\end{equation}
where $\rho^{2}=x^{2}+y^{2}$ and $\hat L_{z}$ is the $z$-component of the electron orbital momentum, which is a constant of motion. In practice  , we consider
the charges $Z_1=3$ and $Z_2=1,2,3$  for Lithium-containing ions.

For the case of unequally charged centers  the only integral of motion apart from the energy is the angular momentum projection $\hat L_z$ on the magnetic field direction, and the eigenstates are labeled by a quantum number corresponding to the excitation, and a Greek letter $\sigma, \pi, \delta$ corresponding to the magnetic quantum number $m=0,-1,-2$, respectively.
In the equilibrium  configuration the eigenstates of the Hamiltonian (\ref{ham-LiX}) with equal charges ($Z_1=Z_2$)  are characterized by two integrals of motion: $\hat L_z$ and the spatial parity operator $\hat P$ ($\vec{r} \to -\vec{r}$) with eigenvalues $p = \pm 1$. The parallel symmetric configuration is also characterized by the $z$-parity,
$\hat P_z (z \to -z)$ with eigenvalues $\sigma=\pm 1$. The magnetic quantum number $m$, spatial parity $p$ and $z$-parity $\sigma$ are related by
\[
   p = \sigma (-1)^{m}\ .
\]
Thus, any eigenstate has two definite quantum numbers: $m$ and $p$. Therefore the space of eigenstates is split into subspaces (sectors) each of them characterized by definite $m$ and $\sigma$, or $m$ and $p$. Notation involves the subscript $g/u$ (gerade/ungerade) corresponding to positive/negative eigenvalues of the spatial parity operator $\hat P$.

\section{Methods}

\subsection{The variational method}

We first study the ground state of the Hamiltonian (\ref{ham-LiX}) by means of    the variational method. We follow the general recipe for the choice of physically relevant trial functions presented in \cite{Tur}. For the case of unequal charged centers we use a  trial function of  a similar form to the one used to study  the ${\rm HeH}^{2+}$ molecular ion in a strong magnetic field (see \cite{Turbiner:2007}),
\begin{equation}
\Psi_{trial} = \psi_{1} +\psi_{2} +\psi_{3}\ ,
\label{Psi-Z1Z2}
\end{equation}
with
\numparts
\begin{eqnarray}
\label{Psitr1:Z1Z2}
 \psi_{1}&=& A_{1}{\mathrm e}^{-\alpha_{1}r_1}{\mathrm e}^{-\beta_1\frac{B}{4}\rho^{2}} +
              A_{2}{\mathrm e}^{-\alpha_{2}r_2}{\mathrm e}^{-\beta_2\frac{B}{4}\rho^{2}}\,, \\
\label{Psitr2:Z1Z2}
  \psi_{2}&=& A_{3}{\mathrm e}^{-\alpha_{3}r_1-\alpha_{4}r_2} {\mathrm e}^{-\beta_{3}\frac{B}{4}\rho^{2}}\,, \\
\label{Psitr3:Z1Z2}
  \psi_{3}&=& A_{4}{\mathrm e}^{-\alpha_{5}r_1-\alpha_{6}r_2}{\mathrm e}^{-\beta_4\frac{B}{4}\rho^{2}} +
              A_{5}{\mathrm e}^{-\alpha_{7}r_1-\alpha_{8}r_2} {\mathrm e}^{-\beta_5\frac{B}{4}\rho^{2}}\,,
\end{eqnarray}
\endnumparts
where $A_{1\ldots 5}$, $\alpha_{1\ldots 8}$, $\beta_{1\ldots 5}$ are variational parameters. Considering the internuclear distance $R$ as a variational parameter, we have in total $19$ variational parameters (a free normalization of the trial function (\ref{Psi-Z1Z2}) allows us to keep fixed one of the parameters $A_{1\ldots 5}$).

For the case of symmetric systems with equal charges $Z_1=Z_2=Z$ we use a ground state trial function which was successfully used to explore the ${\rm H}_2^+$ and ${\rm He}_2^{3+}$ molecular ions in a strong magnetic field \cite{phrep}.
It has the same form (\ref{Psi-Z1Z2}) with permutationally-symmetric functions (\ref{Psitr1:Z1Z2}-\ref{Psitr3:Z1Z2}),
\numparts
\begin{eqnarray}
\label{psi123-He2}
 \psi_1 &=& A_1{({\mathrm e}^{-\alpha_1 r_1} + {\mathrm e}^{-\alpha_1 r_2})\,{\mathrm e}^{-\beta_{1}\frac{B}{4}\rho^{2}}},\\
 \psi_2 &=& A_2\,{ {\mathrm e}^{-\alpha_{2}(r_1 + r_2)}\,{\mathrm e}^{-\beta_{2}\frac{B}{4} \rho^{2}}}\,, \\
 \psi_3 &=& A_3{({\mathrm e}^{-\alpha_3 r_1 -\alpha_4  r_2} +{\mathrm e}^{-\alpha_4  r_1 -\alpha_3  r_2})\,
{\mathrm e}^{-\beta_{3}\frac{B}{4} \rho^{2}}}\,,
\end{eqnarray}
\endnumparts
where $A_{1\ldots 3}$,  $\alpha_{1\ldots 4}$, $\beta_{1\ldots 3}$ are variational parameters.  Here we end up with ten variational parameters.  The functions $\psi_{1,2}$ describe the coherent and incoherent interactions of the electron with the charged centers, respectively. Obviously, they  are suitable modifications of the celebrated Heitler-London and  Hund-Mulliken ${\rm H}_2^+$-trial functions by multiplication with the lowest Landau orbital. In turn, the function $\psi_3$ is a modified Guillemin-Zener function. It can be also considered as a non-linear superposition of the functions $\psi_{1,2}$.

\subsection{The Lagrange-mesh method}

With the aim to confirm and improve the variational results, we apply the Lagrange-mesh method \cite{h2malla,mesh,hmesh} which is an approximate variational calculation performed with the Lagrange basis and simplified by the use of the corresponding Gauss quadrature. So far, this method gives the most accurate total energy of ${\rm H}_2^+$-ion in a magnetic field. The implementation of this method is made easier with an {\it a priori} knowledge of the equilibrium configuration and the total energy of the system. Thus, the variational results obtained for all studied systems $(li, e)$,  $(li, li, e)$,  $(li, \alpha, e)$ and $(li, p, e)$ can serve as input information for the Lagrange-mesh method.

In order to study a two-center molecular system of charges  $Z_{1}$ and $Z_{2}$ it is adequate to use the spheroidal coordinates $(\xi,\eta,\varphi)$ defined as
\begin{eqnarray}
\xi = \frac{r_{1}+r_{2}}{R} -1\,, \hspace{1cm} \eta= \frac{r_{1}-r_{2}}{R} \,,
\end{eqnarray}
(see figure\ref{fig:lix}) with $\xi\in(0,\infty)$, $\eta\in(-1,1)$ and the azimuthal 
angle $\varphi \in (0,2\pi)$. Since the $z$-projection of the orbital angular momentum 
$\hat L_z$ commutes with the Hamiltonian (\ref{ham-LiX}), it is replaced by its eigenvalue $m$ (magnetic quantum number). Because we are interested by the ground state, the wave function should be nodeless, hence $m=0$ and it can be written as~\cite{h2malla}
\begin{equation}
\Psi_{m=0}(\textbf{r}) = \frac{2}{(\pi R^{3})^{1/2}}\psi_{0}(\xi,\eta)\,.
\label{functot}
\end{equation}
Making a substitution of (\ref{functot}) into the  Schr$\ddot{\mathrm o}$dinger 
equation for the Hamiltonian (\ref{ham-LiX}), we arrive at the differential equation for the ground state function $\psi_{0}(\xi,\eta)$,
\begin{equation}
\label{schr0}
\left[\frac{4}{R^{2}J(\xi,\eta)}(T_{\xi}+T_{\eta}) + V(\xi,\eta) \right] \psi_{0}(\xi,\eta) = E \,\psi_{0}(\xi,\eta)\,,
\end{equation}
with the effective potential
\begin{eqnarray}
\fl V(\xi,\eta) =\frac{2}{R}\left(Z_{1}Z_{2}-\frac{2(Z_{1}+Z_{2})(\xi+1)-2\eta(Z_{1}-Z_{2})   }{(\xi+1)^{2}-\eta^{2}}\right) \nonumber \\ + \frac{R^{2}B^{2}}{16}\xi(\xi+2)(1-\eta^{2})\,,
\end{eqnarray}
where
\begin{equation}
\label{jacobian}
J(\xi,\eta)=(\xi+1)^{2}-\eta^{2}\ .
\end{equation}
Here
\begin{equation}
T_{\xi}=-\frac{d}{d\xi}\xi(\xi+2)\frac{d}{d\xi}\,,\hspace{1.5cm}T_{\eta}
=-\frac{d}{d\eta}(1-\eta^2)\frac{d}{d\eta}\ .
\end{equation}
We consider $\psi_{0}(\xi,\eta)$ spanned in some finite basis
\begin{equation}
\label{wavef}
\psi_{0}(\xi,\eta)=\sum_{i=1}^{N_{\xi}}\sum_{j=1}^{N_{\eta}}c_{ij}F_{ij}(\xi,\eta) \ ,
\end{equation}
where $N_{\xi}$ is the size in the $\xi$-direction and $N_{\eta}$ is in the $\eta$-direction, respectively. Explicitly, it is given by (see \cite{h2malla})
\begin{equation}
\label{2dF}
F_{ij}(\xi,\eta)=J_{ij}^{-1/2}f_{i}(\xi)g_{j}(\eta)\,,
\end{equation}
where
\begin{equation}
f_{i}(\xi)=(-1)^{i}(h x_{i})^{1/2}\frac{L_{N_{\xi}}(\xi/h)}{\xi-h x_{i}}{\mathrm e}^{-\xi/2h}\,,
\end{equation}
\begin{equation}
g_{i}(\eta)=(-1)^{N_{\eta}-j}\sqrt{\frac{1-\eta_{j}^{2}}{2}}\frac{P_{N_{\eta}}(\eta)}{\eta-\eta_{j}}\,,
\end{equation}
here $J_{ij}=J(h x_i,\eta_j)$ (see (\ref{jacobian})), where $x_{i}$ and $\eta_{j}$ are the $i$th zero of the Laguerre polynomial $L_{N_{\xi}}(x)$ and the $j$th zero of the Legendre polynomial $P_{N_{\eta}}(\eta)$, respectively.
The dimensionless parameter $h$ is introduced for convenience. Making its variation, it allows us to adapt  the Lagrange mesh to the actual size of the molecular ion. Taking the wave function (\ref{wavef}) and using the Gaussian quadratures associated to each coordinate, the equation (\ref{schr0}) gets the form of mesh equations (see \cite{h2malla})
\begin{equation}
\sum_{i'=1}^{N_{\xi}}\sum_{j'=1}^{N_{\eta}}[ T_{iji'j'}+V(hx_{i},\eta_{j})\,\delta_{ij}\delta_{i'j'} -E\, \delta_{ii'}\delta_{jj'} ]c_{i'j'}=0\, .
\label{matrixE}
\end{equation}
The kinetic energy matrix elements $T_{iji'j'}$ are calculated in \cite{h2malla}. The potential  $V(\xi, \eta)$ is evaluated at the zeros of the Laguerre (scaled by the dimensionless parameter $h$) and Legendre polynomials. Finally, the problem of solving the 
Schr$\ddot{\mathrm o}$dinger equation is reduced to searching eigenvalues of the matrix equation (\ref{matrixE}).

\section{Results}

The results of the variational and mesh calculations for the $m=0$ ground state of the Lithium-containing one-electron Coulomb systems  $(li, e)$ as well as  $(li, li, e)$,  $(li, \alpha, e)$ and $(li, p, e)$  in  magnetic fields $B \le 10^{7}\,$a.u.   in parallel configuration are presented in  tables \ref{lit:atom}, \ref{Tlili}, \ref{Tlihe}, and \ref{Tlih}, respectively.
The results for the molecular systems, show that  the corresponding potential energy curve starts to display a well pronounced minimum at finite internuclear distance at the threshold magnetic fields $B_{th} \sim 2 \times 10^4\, , \sim 10^5\,$ and $\sim 10^6$
\ a.u. It provides a theoretical indication to the possible existence of the bound exotic diatomic molecular ions ${\rm Li}_2^{5+}$,  ${\rm LiHe}^{4+}$, and ${\rm LiH}^{3+}$. For each molecular system, the potential energy curve as a function of the internuclear distance $R$ is characterized by the presence of a potential barrier. At large internuclear distances the interaction of two charged centers becomes repulsive: the total energy curve approaches from above to the total energy of the atomic ion ${\rm Li}^{2+}$ (see Figs.~\ref{potliliB1e06}, \ref{potliheB1e06}, \ref{potlihB5e06}).  If the energy of ${\rm Li}^{2+}$ is lower than the minimum on the potential curve, the system is metastable towards the decay to ${\rm Li}^{2+}$.  All studied systems display two general properties of the Coulomb systems in a magnetic field: as the strength of the magnetic  field increases, they become more bound (the binding energy grows) and more compact (the equilibrium distance decreases).

In tables \ref{Tlili}, \ref{Tlihe}, and \ref{Tlih} we present the  equilibrium internuclear distance $R_{eq}$, the corresponding  total  $E_{t}^{min}$  and binding  $E_{b}= B - E_{t}^{min}$ energies, the dissociation energy $E_{diss}=E_{t}^{min} - E_{t}^{Li^{2+}}$ (corresponding to the decay into  ${\rm Li}^{2+}$), as well as  the position $R_{max}$ of the maximum $E_{t}^{max}$ of the barrier  and its height  {\it i.e.} the difference $\Delta E =E_{t}^{max}-E_{t}^{min}$ for the molecular ions ${\rm Li}_2^{5+}$,  ${\rm LiHe}^{4+}$ and ${\rm LiH}^{3+}$, respectively.

For each molecular system we calculated the lowest vibrational energy $E_{0}^{vib}$ by using the harmonic oscillator approximation around the equilibrium position. It defines the zero point energy and is presented in tables \ref{Tlili}, \ref{Tlihe} and \ref{Tlih}. Finally, $R_{\rm cross}$ indicates the value of the internuclear distance $R$ beyond the maximum for which the total energy is equal to the energy value at the minimum. $R_{eq}$, $R_{max}$, $R_{cross}$ and $\Delta E$ give a qualitative description of the barrier.

Variational calculations were performed in MacBook with an Intel Core 2 Duo CPU at 2.4 GHz.  The minimization package MINUIT from CERN-LIB and an adaptive multidimensional numerical integration routine D01FCF from NAG-LIB were used in the variational calculations. For the lowest-eigenvalue search in the mesh calculations the software code JADAMILU~\cite{JDM} was used.

\subsection{The atomic ion ${\rm Li}^{2+}$}

The hydrogenic ion  ${\rm Li}^{2+}$ exists for all magnetic fields and is stable.
In table~\ref{lit:atom} we present the results obtained for the ground state total energy of ${\rm Li}^{2+}$ in a magnetic field in the range $2.1\times 10^{4}\, \leq B \leq 10^{7}\,$  using both the variational method and the Lagrange-mesh method.
In the variational approach, the trial function is taken as a linear combination of the products of the lowest Coulomb orbital by the lowest Landau orbital
\begin{equation}
\Psi_{trial} = \sum^{4}_{i=1} A_{i}\,{\mathrm e}^{-\alpha_i r-\beta_i \frac{B}{4}\rho^{2}}\,,
\end{equation}
where $A_{1\dots4}$, $\alpha_{1\dots 4}$, $\beta_{1\dots 4}$ are variational parameters. This few parametric trial function provides highly accurate results for the total energy in comparison to the more accurate Lagrange mesh calculations (when converged) for the whole domain of magnetic fields studied (see table~\ref{lit:atom}). They are systematically higher than those obtained with the Lagrange mesh method. Their absolute accuracies increase from $\sim$~0.01 Ry at $B=2.1\times 10^4$\ a.u. to $\sim$~0.04 Ry at $B= 10^6$\ a.u. In turn, the relative accuracy on the binding energies varies from 0.005\% to 0.01\%. A feature that arises here (and which is also found in other systems, see below) is the fact that when we increase the magnetic field, the number of points that are necessary in the Lagrange-mesh method for a reliable (converged) calculation gradually increases as a magnetic field increases. Eventually, it limits the domain of application of the Lagrange-mesh method. For example, with available computer resources it was impossible to get a number of mesh points at $B=5 \times 10^{6}$\ and higher magnetic fields in order to reach convergence.

\begin{table}[!htbp]
\caption{\label{lit:atom} Results for the total ($E_{t}$) and binding ($E_{b}$) energies (in Ry) of the $1s_0$ ground state of the atomic ion ${\rm Li}^{2+}$ in a magnetic field using the variational method (upper lines) and the Lagrange-mesh method (lower lines) when available.}
\lineup
\begin{indented}
\item[]\begin{tabular}{@{}l l l }
\br
$B$ & $E_{t}$ & $E_{b}$ \\
\mr
\multirow{2}{*}{$2.1\times10^{4}$}
&\0\0\0\020\,825.331 &\0\0174.6695\\
&\0\0\0\020\,825.3232&\0\0174.6768\\
\bs
\multirow{2}{*}{$5\times10^{4}$}
&\0\0\0\049\,780.425 &\0\0219.5746\\
&\0\0\0\049\,780.4084&\0\0219.5916\\
\bs
\multirow{2}{*}{$10^{5}$}
&\0\0\0\099\,738.828 &\0\0261.1723\\
&\0\0\0\099\,738.7958&\0\0261.2042\\
\bs
\multirow{2}{*}{$5\times10^{5}$}
&\0\0\0499\,620.917&\0\0379.083\\
&\0\0\0499\,620.883&\0\0379.117\\
\bs
\multirow{2}{*}{$10^{6}$}
&\0\0\0999\,560.377&\0\0439.623\\
&\0\0\0999\,560.340&\0\0439.660\\
\bs
\multirow{2}{*}{$5\times10^{6}$}
&\0\04\,999\,395.71&\0\0604.29   \\
&$<4\,999\,396.4$  &$>603.6$ \\
\bs
\multirow{1}{*}{$10^{7}$}
&\0\09\,999\,314.04&\0\0685.96\\
\br
\end{tabular}
\end{indented}
\end{table}

\subsection{The molecular ion ${\rm Li}_{2}^{5+}$}

\begin{table}[!htbp]
\caption{\label{Tlili} Molecular ion ${\rm Li}_{2}^{5+}$ in a magnetic field $B$:
equilibrium internuclear distance $R_{eq}$, total $E_{t}^{min}$ and binding $E_{b} = B-E_{t}^{min}$ energies in the variational method (upper lines) and the Lagrange-mesh method (lower lines); dissociation energy $E_{diss}=E_{t}^{min}-E_{t}^{\rm Li^{2+}}$; $R_{max}$ of the maximum at the potential curve (see figure~\ref{potliliB1e06}); height of the barrier $\Delta E=E_{t}^{max}-E_{t}^{min}$ and lowest vibrational energy $E_0^{vib}$; $R_{cross}$ is the distance for which $E_{t}=E_{t}^{min}$.}
\lineup
\begin{tabular}{@{}lllllllll}
\br
$B$& $R_{eq}$ & $E_{t}^{min}$ &  $E_{b}$ &  $E_{diss}$ & $R_{max}$ & $\Delta E$ & $E_{0}^{vib}$  & $R_{cross}$ \\
\mr
\multirow{2}{*}{$2.1\times10^{4}$}
 &0.211 &\0\0\,20\,854.933 &145.067 &\m29.603 &0.234&\0\00.014&0.11&0.247\\
 &0.2118&\0\0\,20\,854.8373&145.1627&\m29.5141&0.232&\0\00.0096&0.11 &0.2118 \\
\bs
\multirow{2}{*}{$5\times10^{4}$}
 &0.139 &\0\0\,49\,809.483 &190.517 &\m29.058&0.252&\0\02.90  &0.48&0.37\\
 &0.1394&\0\0\,49\,809.3904&190.6096&\m28.9820&0.252&\0\02.8843&0.48&0.3655 \\
\bs
\multirow{2}{*}{$10^{5}$}
 &0.111 &\0\0\,99\,764.908 &235.092 &\m26.080 &0.245&\0\07.99&0.76&0.45\\
 &0.1107&\0\0\,99\,764.8113&235.1887&\m26.0155&0.246&\0\07.98&0.76&0.4451\\
\bs
\multirow{2}{*}{$5\times10^{5}$}
 &0.071 &\0\,499\,628.776&371.224&\0\m7.859&0.225 &\031.51&1.69&1.56 \\
 &0.0707&\0\,499\,628.631&371.369&\0\m7.748&0.2246&\031.6 &1.69&\dots\\
\bs
\multirow{2}{*}{$10^{6}$}
 &0.060 &\0\,999\,554.225&445.775&\0$-6.15$&0.215&\048.06&2.33& \\
 &0.0597&\0\,999\,554.020&445.980&\0$-6.32$&$\dots$&$\dots$&2.26& \\
\bs
\multirow{1}{*}{$10^{7}$}
 &0.037&9\,999\,225.689& 774.311&$-88.35$&0.186&138.59&5.98& \\
\br
\end{tabular}
\end{table}

\begin{figure}[!hbtp]
\centering
\includegraphics[width=0.7\textwidth]{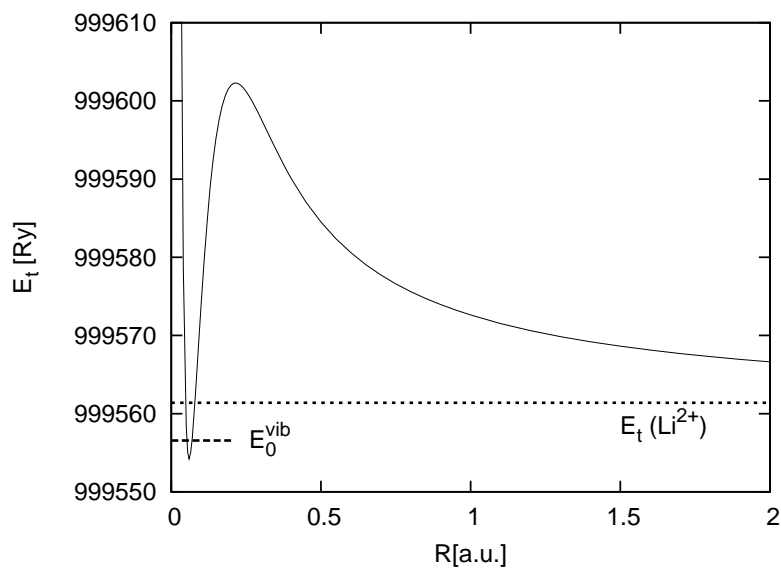}
\caption{\label{potliliB1e06} Total energy for the $1\sigma_g$ state of the ${\rm Li}_{2}^{5+}$ in parallel configuration at magnetic field $B=10^{6}$\ a.u.  as a function of the internuclear distance $R$.  The ${\rm Li}^{2+}$ energy (dotted line) and the lowest vibrational energy (dashed line) are displayed. At the minimum $R_{eq}\approx 0.06$\ a.u. }
\end{figure}

The results of the variational and the Lagrange mesh calculations for the
$1\sigma_g$ state of the one-electron system containing two Lithium nuclei,
$(li, li, e^-)$ for magnetic fields $B\le 10^{7}$\ a.u. in parallel configuration are presented in table~\ref{Tlili}.
It is worth mentioning that for one-electron symmetric systems
with two nuclei with charge $Z$ at distance $R$, the total energy is
related to the total energy of the H$_2^+$ by a relation similar
to a scaling relation for the single nucleus case
presented in \cite{SO-74},
\begin{equation}
E_t (Z,B,R) = Z^2 E_t (1,B/Z^2,ZR) + \frac{2Z(Z-1)}{R}\ .
\end{equation}
Though it is known that H$_2^+$ is stable at all studied magnetic fields
$0 < B < 10^7$\ a.u. (see~\cite{chains2009}), the presence in the r.h.s. of the repulsive
term $2Z(Z-1)/R$ (which is absent in the single nucleus case) indicates
that other one-electron molecular ions may not exist.
For example, for $Z=2$ the He$_2^{3+}$ ion does not exist at $B <
10^2$\ a.u. but it starts to exist for $B \gtrsim 10^2$\ a.u. at first
as a metastable state and then for $B \gtrsim 10^4$\ a.u. as a stable
one. For $Z=3$
the ${\rm Li}_{2}^{5+}$ ion does not exist for $B \lesssim 2.1 \times
10^{4}$\ a.u. (see the discussion below).

A trial function of form (\ref{Psi-Z1Z2}) is used in the variational calculations. For each value of the magnetic field shown in table~\ref{Tlili}, the upper line corresponds to the variational results while the lower line (when displayed) corresponds to the Lagrange-mesh results. For $B=10^{7}$\ a.u., it was impossible to get a sufficiently large number of mesh points with our available computer resources in order to obtain a converged result. The converged Lagrange-mesh energies when available are better than the corresponding variational results. The absolute accuracies of the variational energies increase from about $\sim$ 0.1\ Ry for $B \le 10^5$\ a.u. to $\sim$ 0.2 Ry at $B = 10^6$\ a.u. For the binding energy, the relative accuracy is $\sim 0.07\%$ at $B=2.1 \times 10^{4}$\ a.u. while it is $\sim 0.05\%$ at $B=10^6$\ a.u.

For a magnetic field $B \sim 2.1 \times 10^{4}$\ a.u. the potential energy curve starts to display a  minimum for $R=0.21$\ a.u. indicating the existence of a metastable state of the molecular ion ${\rm Li}_{2}^{5+}$, unstable towards decay to ${\rm Li}_{2}^{5+} \rightarrow {\rm Li}^{2+} +\, li$. Further calculations of the energy curve show that the potential barrier is not high enough to keep a vibrational level, $E_0^{vib} > \Delta E$. Hence, the system is unstable against vibrations. For magnetic fields $B \gtrsim 5\times 10^{4}$\ a.u., the potential energy well becomes sufficiently deep to keep at least one vibrational state. The height of the barrier increases when the magnetic field is increased and the system becomes more stable against vibrations, {\it i.e.}, while both $\Delta E$ and $E_0^{vib}$ increase monotonously with the magnetic field, their ratio   $\Delta E/E_0^{vib}$ increases from $\sim 6$ for $B = 5\times 10^4$\ a.u. up to $\sim 23$ for $B=10^7$\ a.u. At the same time, the dissociation energy $E_{diss}=E_{t}^{min}-E_{t}^{\rm Li^{2+}}$ decreases and eventually, for magnetic fields $B \sim 10^{6}$\ a.u., the total energy becomes smaller than the total energy of the ${\rm Li}^{2+}$ atomic ion. The system becomes stable towards the decay ${\rm Li}_{2}^{5+} \rightarrow {\rm Li}^{2+} +\,li$, a situation depicted in figure~\ref{potliliB1e06} for $B=10^6$\ a.u. The qualitative behaviour of this molecular system is typical: when we increase the magnetic field strength, the molecular ion ${\rm Li}_{2}^{5+}$ becomes more compact (the equilibrium distance decreases) and more bound (the binding energy increases).

\begin{figure}[!htp]
\centering
\includegraphics[width=0.42\textwidth]{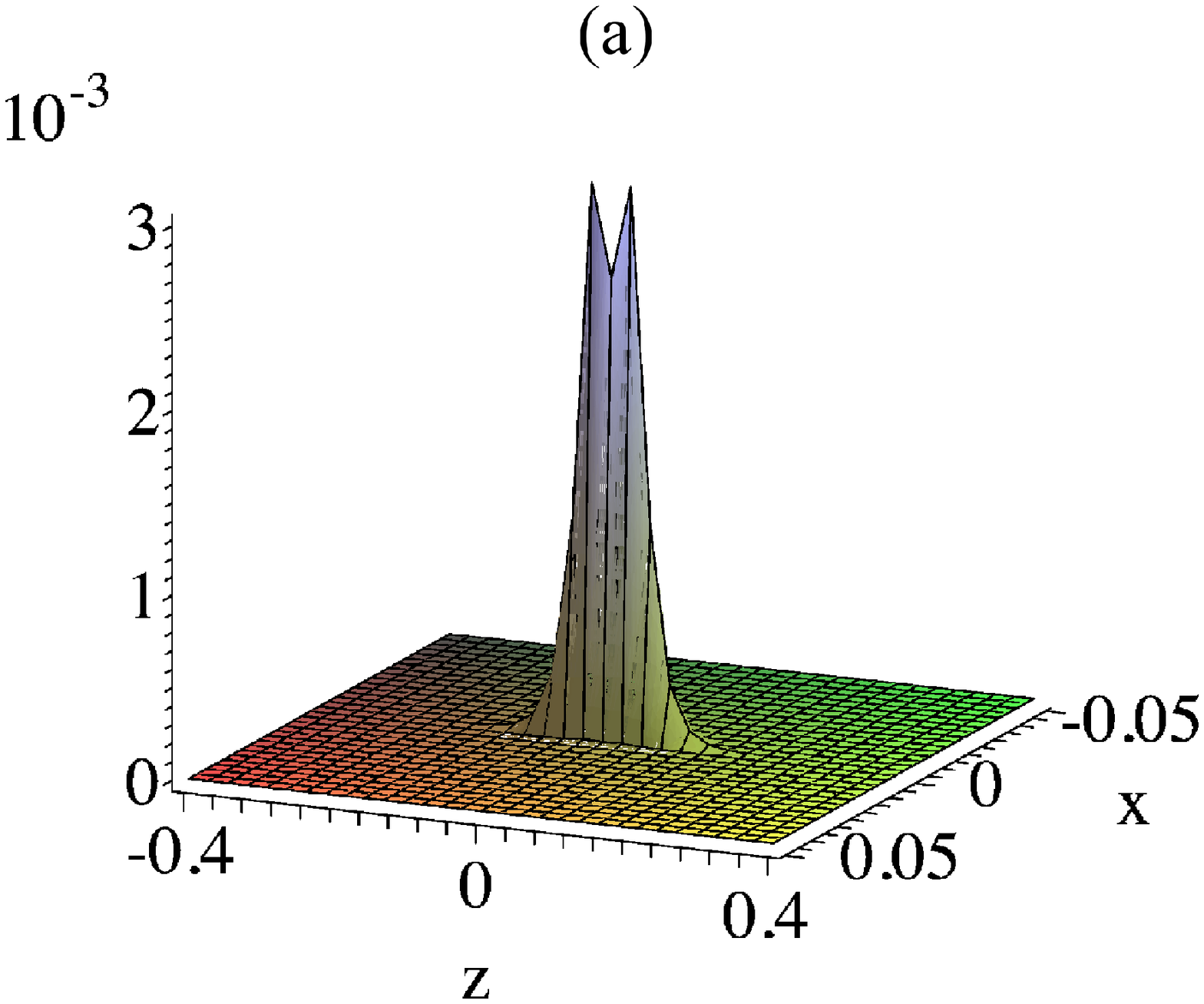}\\
\includegraphics[width=0.42\textwidth]{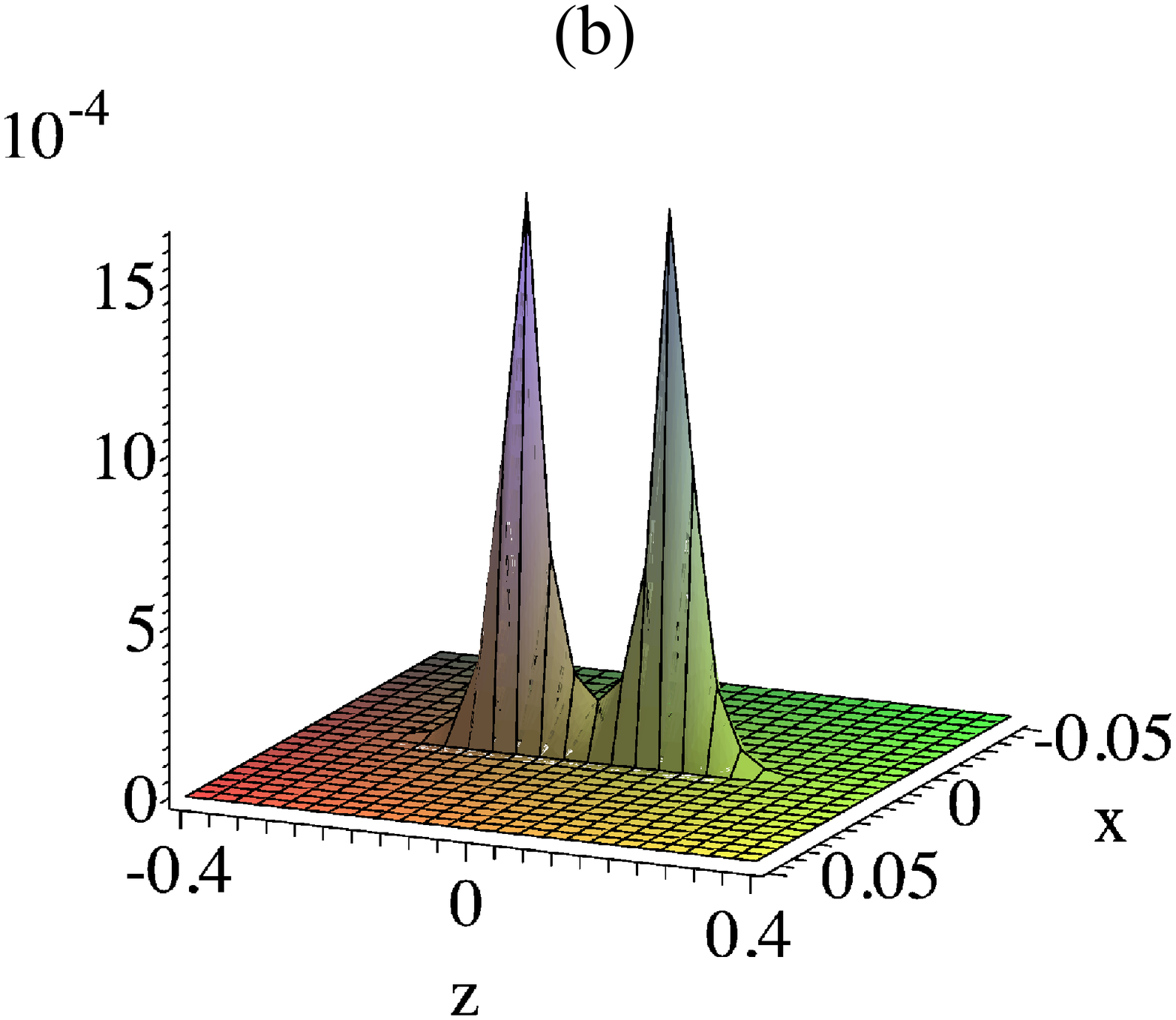}\\
\includegraphics[width=0.42\textwidth]{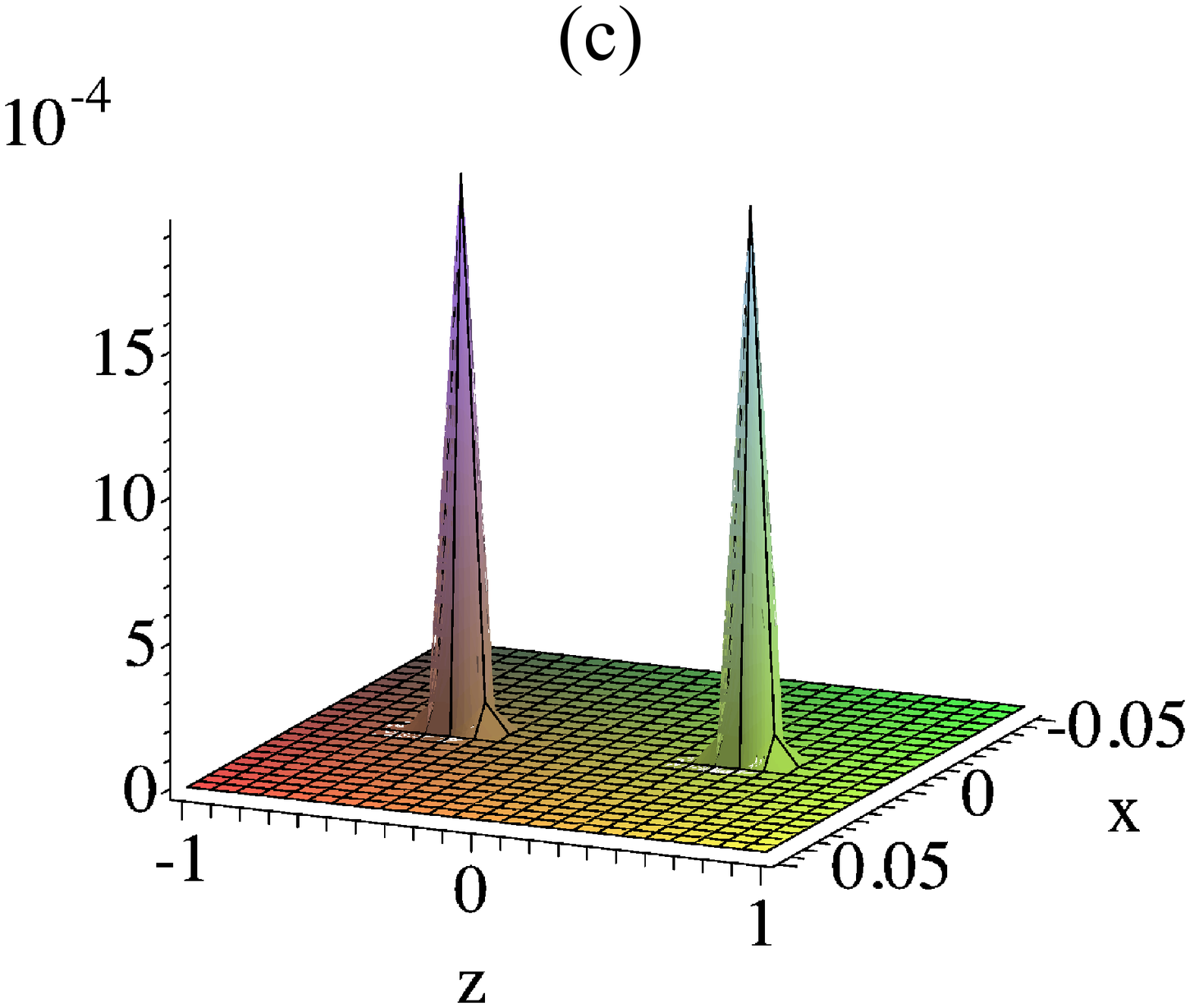}\\
\caption{Evolution of the electronic distribution $\int |\psi_0(x,y,z)|^2 dy$  for the ground state $1\sigma_g$ of ${\rm Li}_{2}^{5+}$ in a magnetic field $B=10^{6}$~a.u. for internuclear distances: (a) $R_{eq}\approx0.06$, (b) $R_{max}\approx0.21$ and (c) $R= 1.0$~a.u. }
\label{EDliliB1e06Rmino}
\end{figure}

The electronic distribution of the ${\rm Li}_{2}^{5+}$ molecular ion in the $1\sigma_g$ state for \hbox{$B=10^{6}$\ a.u.  } is displayed in figure~\ref{EDliliB1e06Rmino} for the  internuclear distances $R_{eq}\approx 0.06$, $R_{max}\approx0.21$, and $R=1$\ a.u. At equilibrium, the electronic distribution is characterized by two overlapping peaks centered at the positions of the charged centers. Eventually, at large internuclear distances the electronic distribution consists of two symmetric electronic distributions each of them corresponding to the atomic ion ${\rm Li}^{2+}$ in a magnetic field. The schematic picture shown in figure~\ref{EDliliB1e06Rmino} for the ground state electronic distribution of ${\rm Li}_{2}^{5+}$  is typical for all the magnetic fields where this molecular ion exists.

\subsection{The molecular ion ${\rm LiHe}^{4+}$}

\begin{table}[!htp]
\caption{\label{Tlihe} Same as in table \ref{Tlili} for the molecular ion ${\rm LiHe}^{4+}$ in a magnetic field $B$.}
\lineup
\begin{tabular}{@{}lllllllll}
\br
$B$& $R_{eq}$ & $E_{t}^{min}$ &  $E_{b}$ &  $E_{diss}$ & $R_{max}$ & $\Delta E$ & $E_{0}^{vib}$  & $R_{cross}$ \\
\mr
\multirow{2}{*}{$10^{5}$}
&0.110 &\0\0\,99\,775.455 &224.545 &\m36.628 &0.144&\00.40  &0.56&0.1663 \\
&0.1101&\0\0\,99\,775.3783&224.6217&\m36.5825&0.144&\00.3917&0.55 &0.1661\\
\bs
\multirow{2}{*}{$5\times10^{5}$}
&0.067&\0\,499\,654.950&345.050&\m34.033&0.136&\09.229&1.72&0.2268\\
&0.067&\0\,499\,654.857&345.142&\m33.974&0.136&\09.213&1.57&0.2267\\
\bs
\multirow{2}{*}{$10^{6}$}
&0.057 &\0\,999\,589.934&410.066&\m29.557&0.130&16.630&2.31&0.2679\\
&0.0568&\0\,999\,589.810&410.190&\m29.470&0.130&16.610&2.30&$\dots$ \\
\bs
\multirow{1}{*}{$5\times10^{6}$}
&0.040  &4\,999\,404.74&595.26 &\0\m9.02 &0.117 &44.40  &4.26   &0.940\\
\bs
\multirow{1}{*}{$10^{7}$}
&0.035&9\,999\,309.05&690.95&\0$-4.99$&0.111&61.42&5.35&  \\
\br
\end{tabular}
\end{table}

Table~\ref{Tlihe} contains the results of the variational and the Lagrange-mesh calculations for the $1\sigma$ state of the one-electron  system  $(li,\alpha, e)$,  for magnetic fields $B \le 10^{7}$\ a.u.  in parallel configuration. For this asymmetric system, a trial function of the form  (\ref{Psi-Z1Z2}) is used in the variational calculations.
For each value of the magnetic field shown in table~\ref{Tlihe}, the upper line corresponds to the variational results while the lower line (when displayed) corresponds to the Lagrange-mesh results. For the binding energy, the relative improvement of the mesh calculations is $\sim 0.03\%$, almost uniformly for magnetic fields  $B= 10^{5}- 10^{6}$~a.u. For $B=5\times 10^{6}$\ a.u. the Lagrange-mesh energy is worse than the variational one.

\begin{figure}[!htp]
\centering
\includegraphics[width=0.7\textwidth]{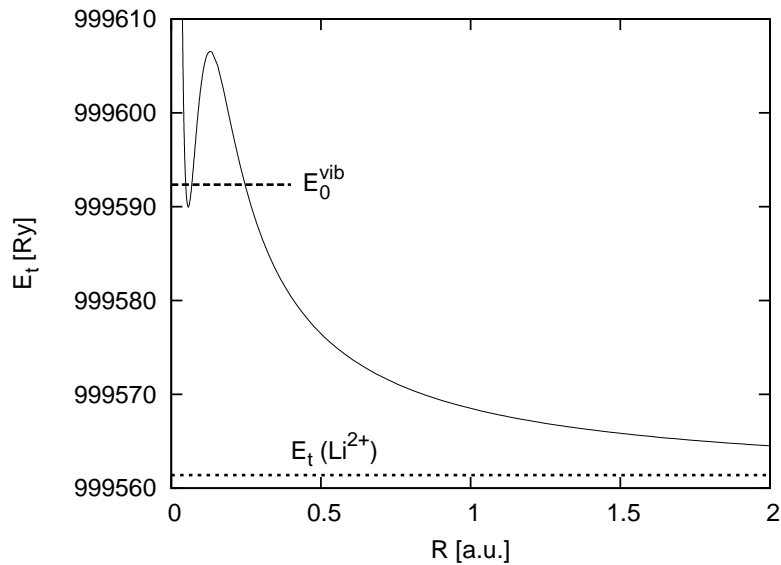}
\caption{\label{potliheB1e06} Total energy for the ground state $1\sigma$ of the molecular ion ${\rm LiHe}^{4+}$ placed in a magnetic field $B=10^{6}$\ a.u. in parallel configuration. The ${\rm Li}^{2+}$ energy (dotted line) and the lowest vibrational energy (dashed line) are displayed. The equilibrium distance $R_{eq}\approx 0.057$\ a.u.}
\end{figure}

For a magnetic field $B\sim 10^{5}$\ a.u., the  potential energy curve  starts to display a  minimum for $R\approx0.11$\ a.u indicating the formation of a metastable state, unstable for the decay to ${\rm LiHe}^{4+} \rightarrow {\rm Li}^{2+} + \alpha$. Also the potential barrier is small and it does not allow to keep a vibrational level $E_0^{vib} > \Delta E$. For magnetic fields $B \gtrsim 5\times 10^{5}$\ a.u., the potential energy well is sufficiently deep to keep more than one vibrational state. Also, the height of the barrier increases when the magnetic field is increased and the system becomes more stable against vibrations,  {\it i.e.} the ratio $\Delta E/E_0^{vib}$ increases from $\sim 5$ for $B=5\times 10^5$\ a.u. up to $\sim 12$ for $B=10^7$\ a.u.
At the same time, the dissociation energy $E_{diss}=E_{t}^{min}-E_{t}^{\rm Li^{2+}}$ decreases and, eventually, for magnetic fields $B\sim 10^{7}$\ a.u., the total energy becomes smaller than the total energy of the atomic ion ${\rm Li}^{2+}$, indicating that the system can become stable towards the decay ${\rm LiHe}^{4+}\rightarrow {\rm Li}^{2+} + \alpha$. However, even at $B = 10^7$\ a.u., the system is still probably metastable with respect to vibrations since $E_0^{vib} > E_{diss}$.
The total energy $E_{t}$ as a function of the internuclear distance $R$ for $B=10^{6}$\ a.u., is shown in figure~\ref{potliheB1e06}.

\begin{figure}[!htp]
\centering
\includegraphics[width=0.42\textwidth]{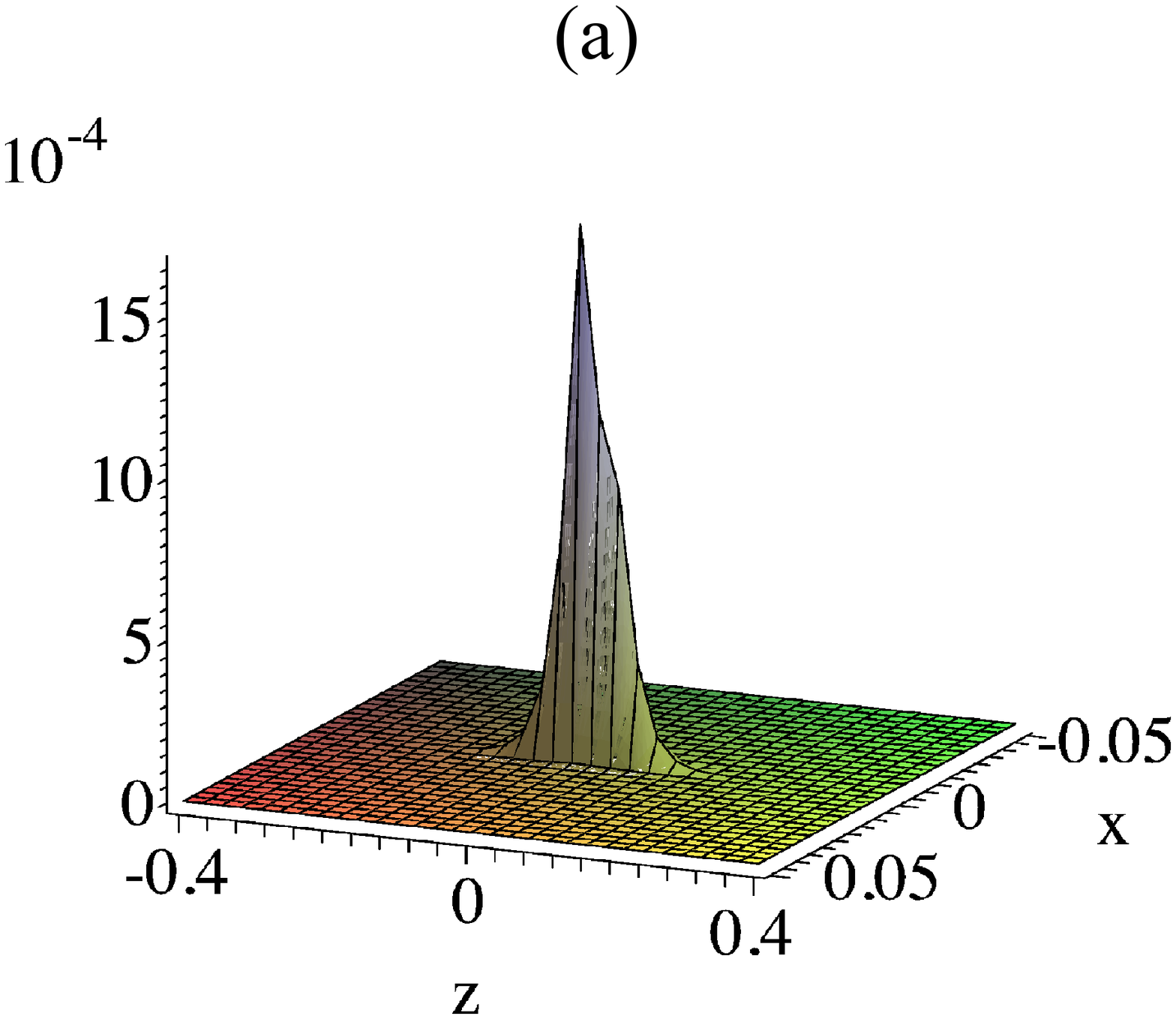}\\
\includegraphics[width=0.42\textwidth]{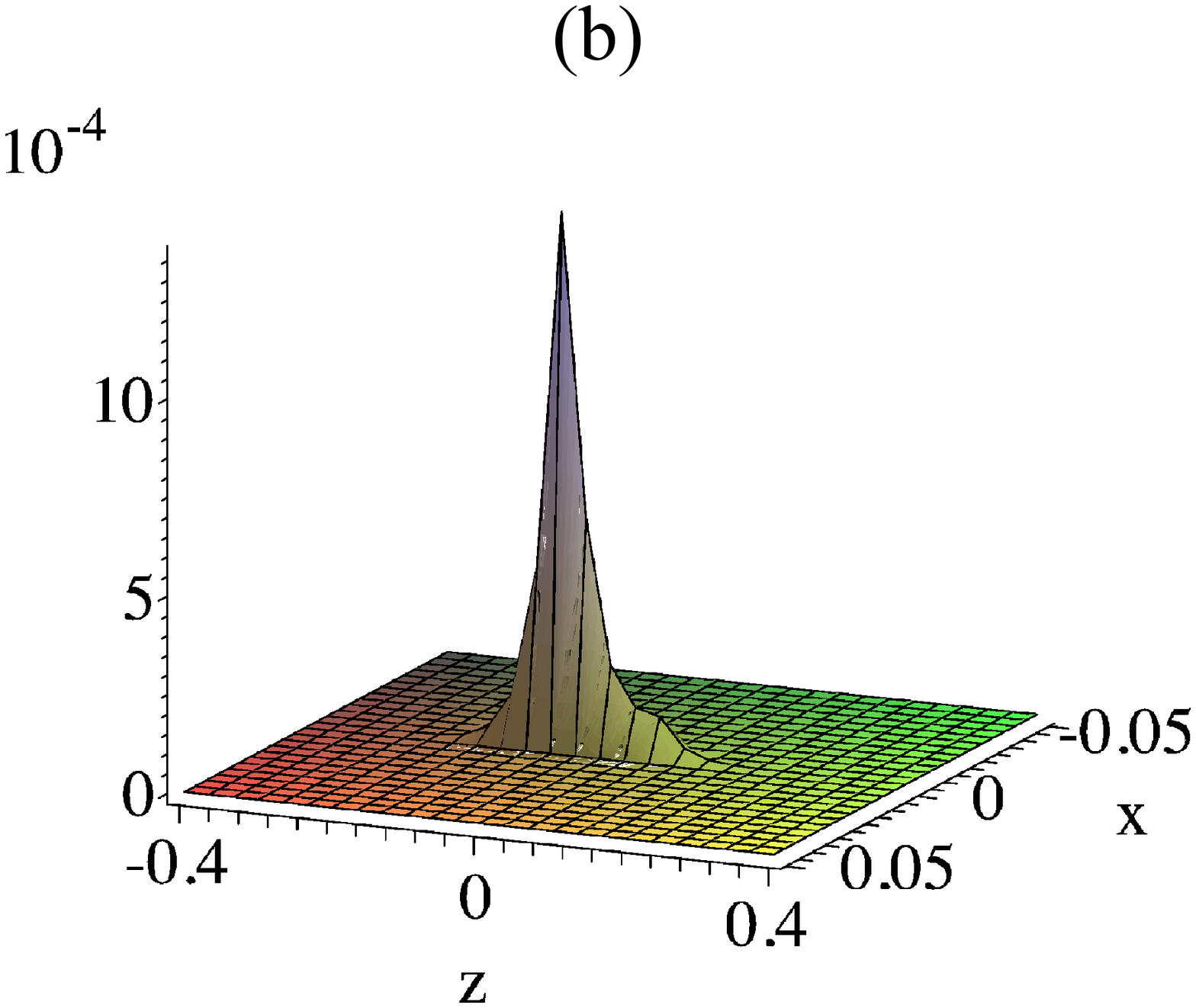}\\
\includegraphics[width=0.42\textwidth]{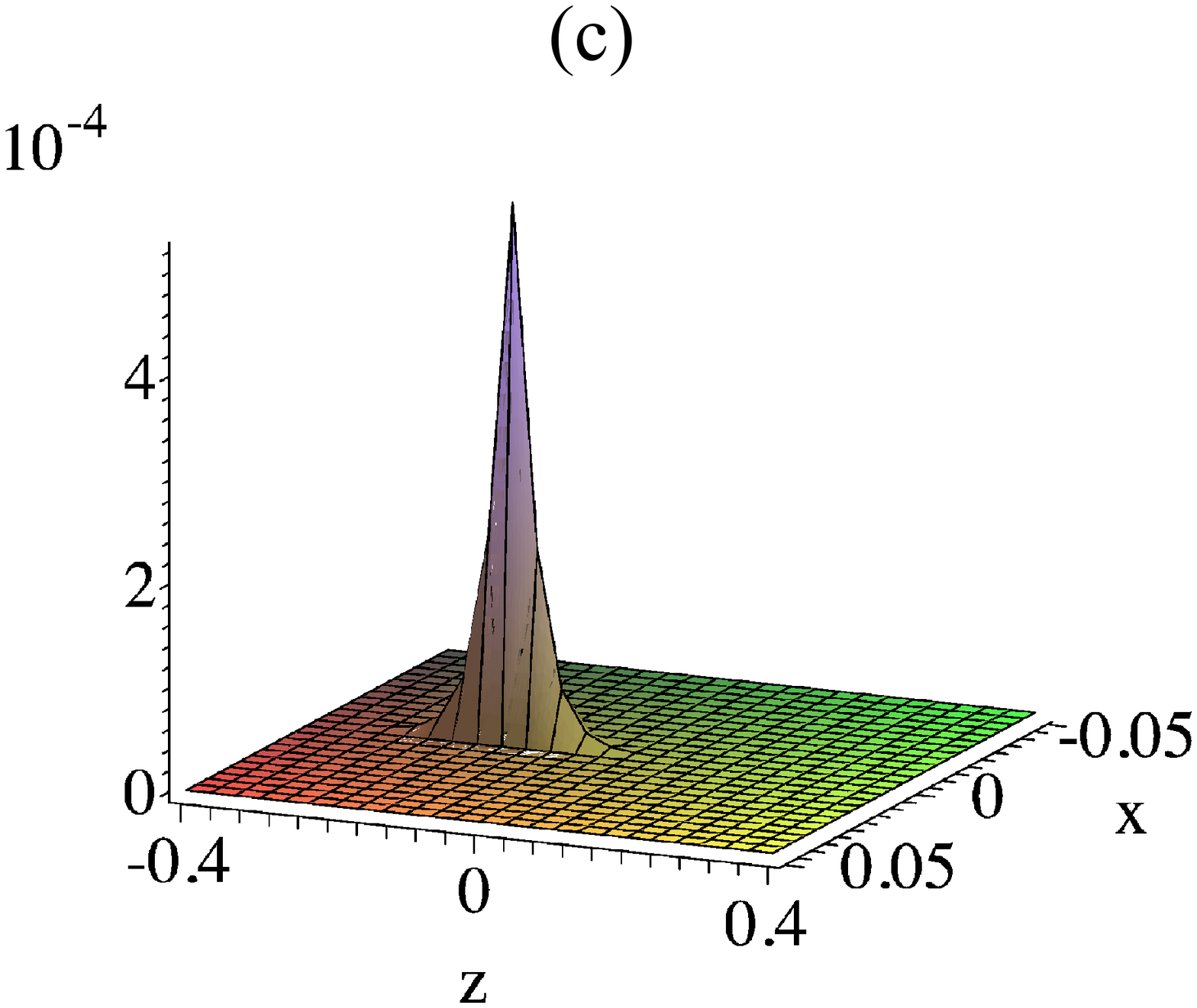}
\caption{Electronic distribution $\int|\psi_{0}(x,y,z)|^{2}dy$ for the molecular ion ${\rm LiHe}^{4+}$ in a magnetic field $B=10^{6}$\ a.u. as a function of the internuclear distance: (a) $R_{eq}\approx0.06$, (b) $R_{max} \approx0.13$  and (c) $R_{cross}\approx  0.27$\ a.u.  }
\label{DE:lihe}
\end{figure}

The electronic distribution of the molecular ion ${\rm LiHe}^{4+}$ in its $1\sigma$ ground state for $B=10^{6}$\ a.u.   is displayed in figure~\ref{DE:lihe} for three different internuclear distances: $R_{eq}~\approx~0.06$, $R_{max}~\approx~0.13$  and $R_{cross}~\approx~0.27$\ a.u.
The profile of the electronic distribution is asymmetric. At equilibrium, the electronic distribution for ${\rm LiHe}^{4+}$ is characterized by one pronounced peak at the position of the Li nucleus and a shoulder due to the presence of the $\alpha$ particle. The electronic distribution  evolves in such a way that eventually the shoulder disappears and the distribution becomes centered at the position of the Li nucleus and less asymmetrical. This ionic-bonding behaviour is typical for all the magnetic fields studied.

\begin{figure}[!htp]
\centering
\includegraphics[width=0.41\textwidth]{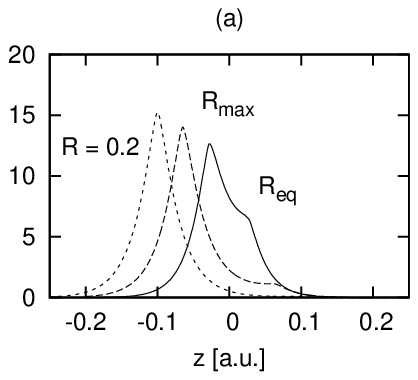}
\includegraphics[width=0.41\textwidth]{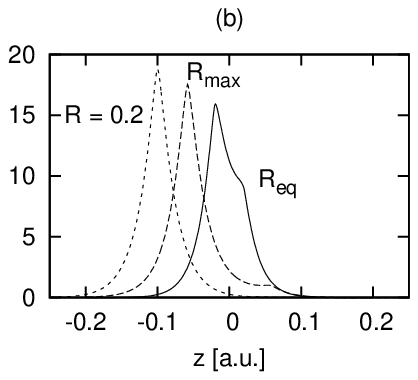}
\caption{$z$-profile of the electronic distribution $\int\int|\psi_{0}(x,y,z)|^{2}dxdy$ of the molecular ion ${\rm LiHe}^{4+}$ for internuclear distances: $R_{eq}$, $R_{max}$, $R=0.2$\ a.u.  and magnetic fields: (a) $B=10^{6}$\ a.u.  and (b) $B=5\times 10^{6}$\ a.u.}
\label{ZDElihe}
\end{figure}

In order to have a better picture of this evolution, figure~\ref{ZDElihe} presents the $z$-profile of the electronic distribution as a function of the internuclear distance for two values of the magnetic field $B=10^{6}$\ a.u.  and $B=5\times10^{6}$\ a.u.  The electronic cloud becomes narrower in the $z$-direction when the magnetic field increases.

\subsection{The molecular ion ${\rm LiH}^{3+}$}

In table~\ref{Tlih} we present the results of the variational and mesh calculations for the $1\sigma$ state of system  $(li, p, e)$ for  magnetic fields $B\le 10^{7}$\ a.u. The trial functions (\ref{Psi-Z1Z2}) are used for this asymmetric system. For a magnetic field $B\sim  10^{6}$\ a.u., the  potential energy curve  starts to display a minimum for $R\approx 0.06$\ a.u. The potential barrier does not keep a vibrational level, $E_0^{vib} > \Delta E$. For $B \gtrsim 5\times 10^{6}$\ a.u. the potential energy well is sufficiently deep to keep at least one vibrational state. However, this system remains unstable towards the decay ${\rm LiH}^{3+}   \rightarrow {\rm Li}^{2+} + p$ for all magnetic fields studied.

\begin{table}[!htp]
\caption{\label{Tlih} Same as table \ref{Tlili} for the molecular ion ${\rm LiH}^{3+}$ in a magnetic field $B$.}
\lineup
\begin{tabular}{@{}lllllllll}
\br
$B$& $R_{eq}$ & $E_{t}^{min}$ &  $E_{b}$ &  $E_{diss}$ & $R_{max}$ & $\Delta E$ & $E_{0}^{vib}$  & $R_{cross}$ \\
\mr
\multirow{2}{*}{$10^{6}$}
 &0.0621 &\0\0\0\,999\,587.579 &\0\0412.421 &27.20 &0.092&\00.95 &1.741 &0.1173\\
 &0.0622 &\0\0\0\,999\,587.42 &\0\0412.57  &27.08 &0.0928&\00.96  &      &0.1175\\
\bs
\multirow{2}{*}{$5\times10^{6}$}
 &0.0415 &\0\04\,999\,419.11&\0\0580.89 &23.39&0.089&\08.4 &4.403 &0.1668 \\
 &0.0415 &$<4\,999\,419.3$&$>580.7$ &$\dots$&$\dots$&$\dots$&$\dots$&$\dots$ \\
\bs
\multirow{1}{*}{$10^{7}$}
  &0.0360 &\0\09\,999\,333.65 &\0\0666.35 &19.61&0.087 &14.2 &5.385 &0.2048\\
\br
\end{tabular}
\end{table}

Only for the magnetic field $B=10^6$\ a.u., we could obtain converged results with the mesh method. For the binding energy,  the relative improvement was $\sim 0.04\%$. For $B=5\times10^{6}$\ a.u., we also present the best Lagrange-mesh results obtained, although they are less accurate than the variational ones.

\begin{figure}[!htp]
\centering
\includegraphics[width=0.7\textwidth]{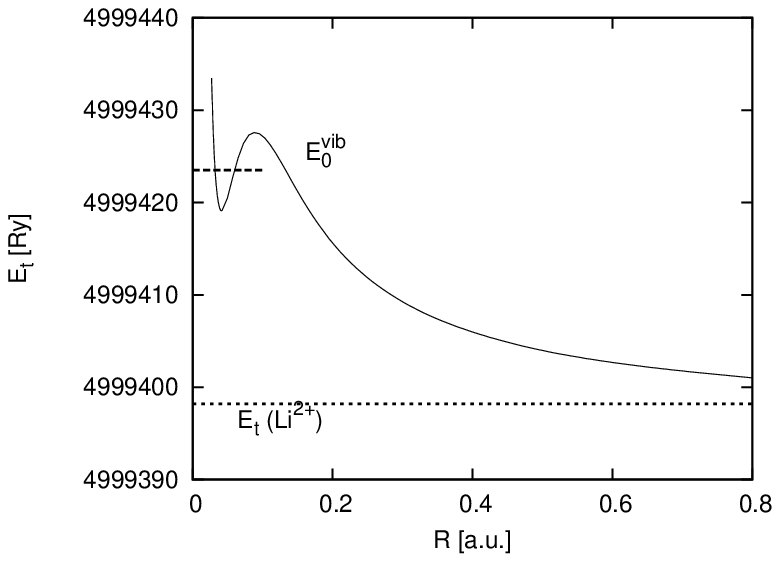}
\caption{Total energy for the $1\sigma$-state of the ${\rm LiH}^{3+}$-ion in parallel configuration at magnetic field $B=5\times10^{6}$\ a.u. as a function of the internuclear distance $R$. The total energy of ${\rm Li}^{2+}$ (dotted line) and the lowest vibrational energy (dashed line) are displayed. The minimum occurs at $R_{eq}\approx 0.0415$\ a.u.  }
\label{potlihB5e06}
\end{figure}

\begin{figure}[!htp]
\centering
\includegraphics[width=0.42\textwidth]{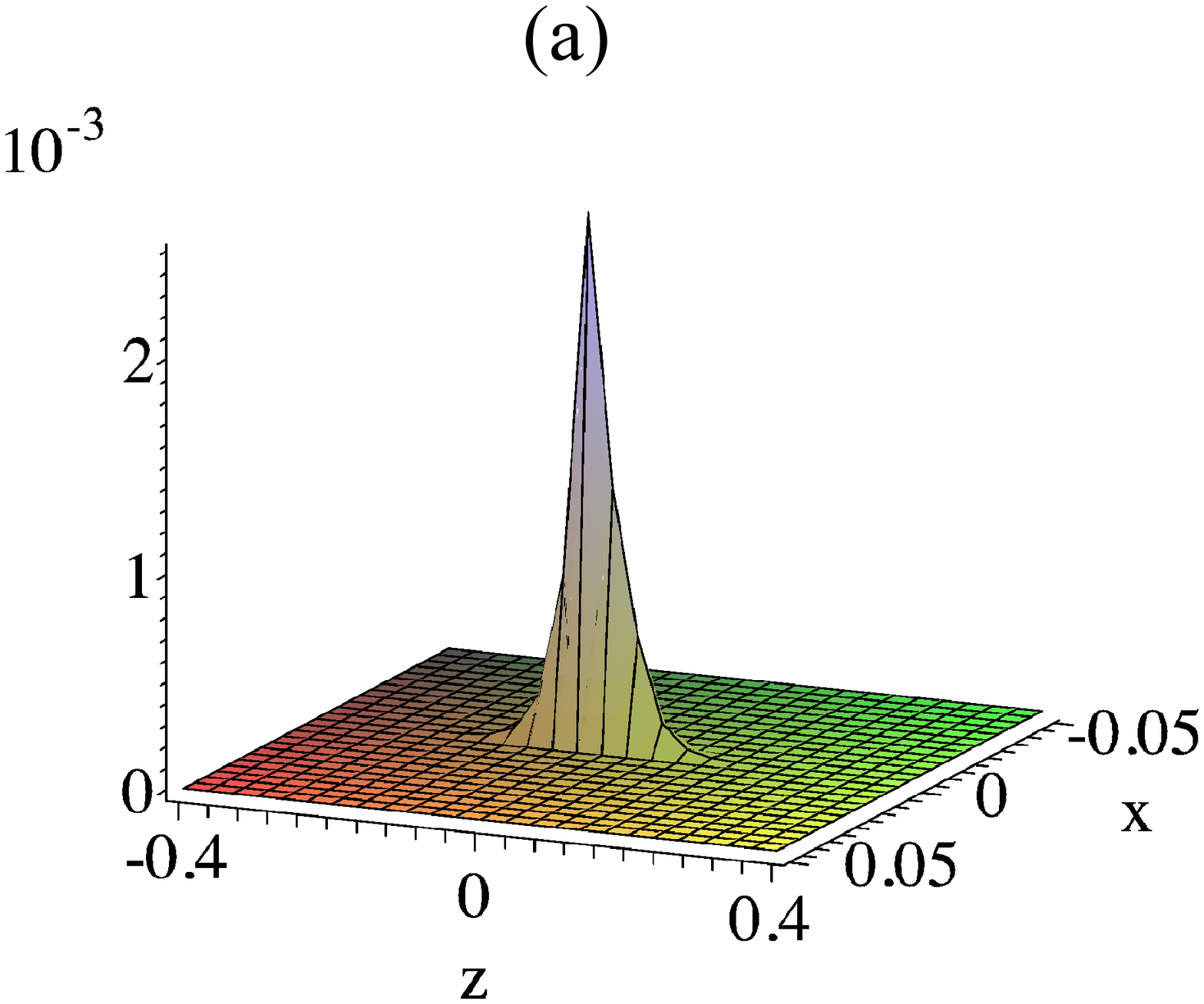}\\
\includegraphics[width=0.42\textwidth]{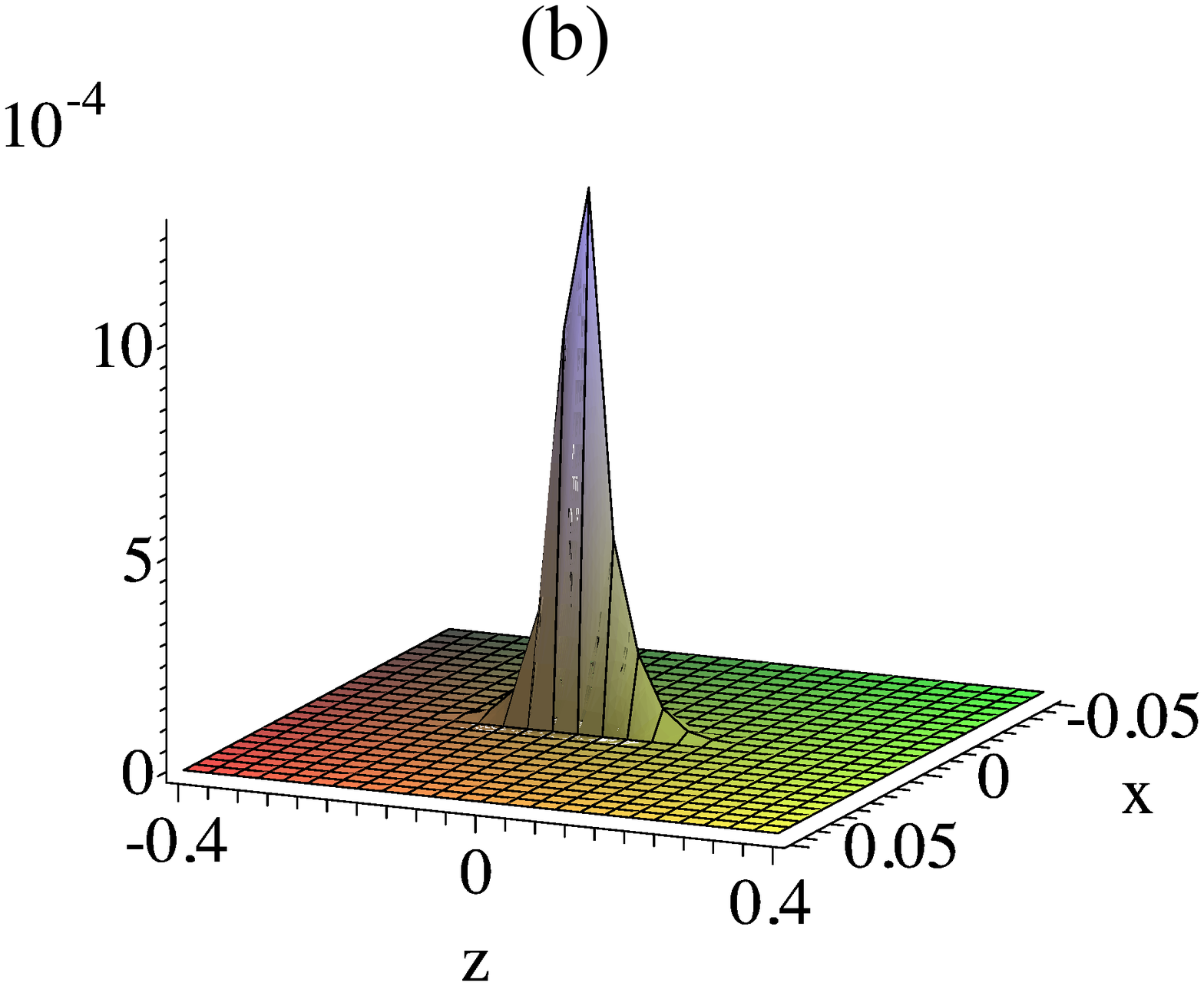}\\
\includegraphics[width=0.42\textwidth]{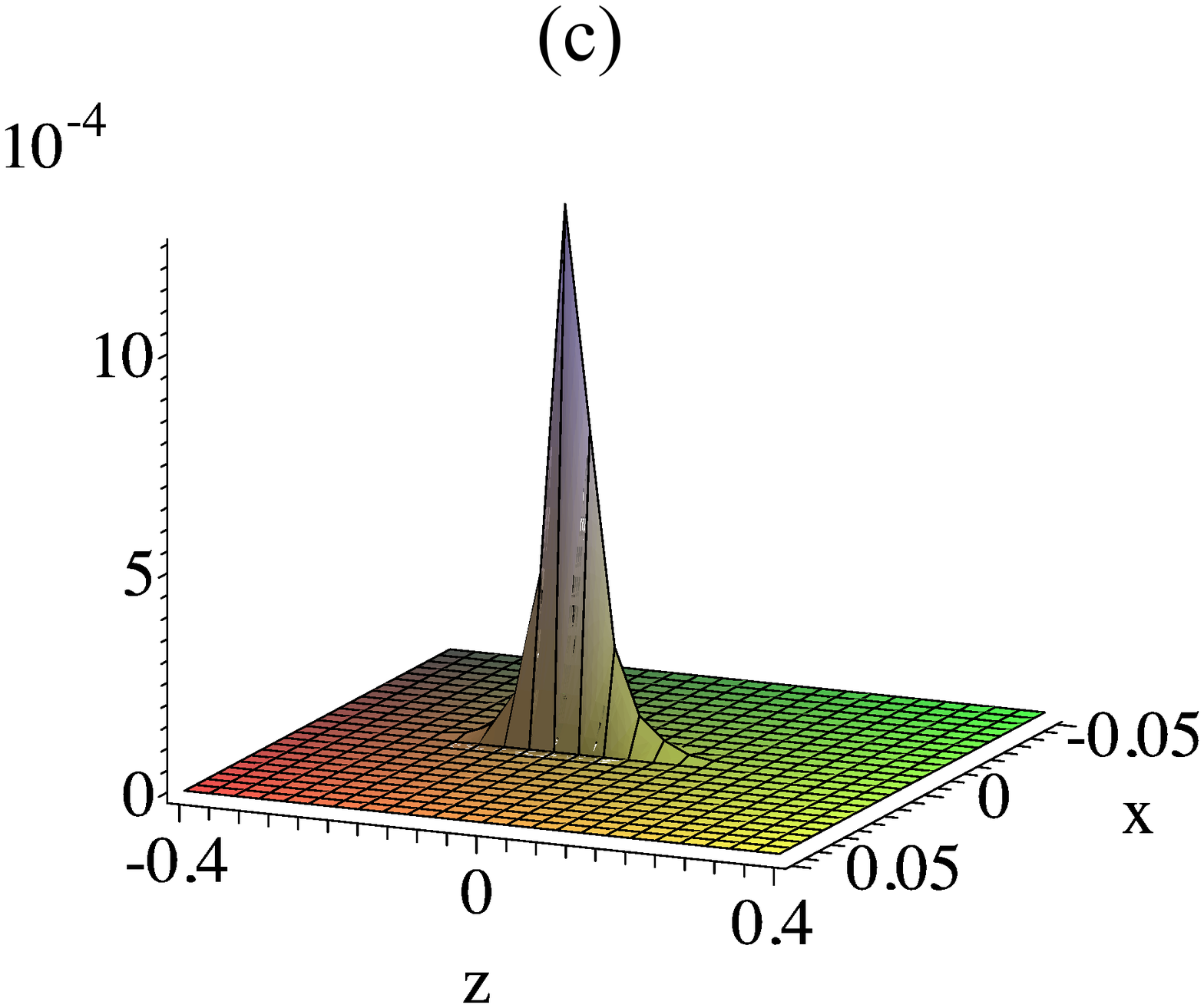}
\caption{Electronic distribution $\int|\psi_{0}(x,y,z)|^{2}dy$ of the ground state the molecular ion ${\rm LiH}^{3+}$ in a magnetic field $B=10^{6}$\ a.u. for internuclear distances:  (a) $R\approx 0.06$ , (b) $R\approx 0.09$  and (c) $R \approx 0.12$\ a.u.}
\label{DE:lih}
\end{figure}

\begin{figure}[!htp]
\centering
\includegraphics[width=0.41\textwidth]{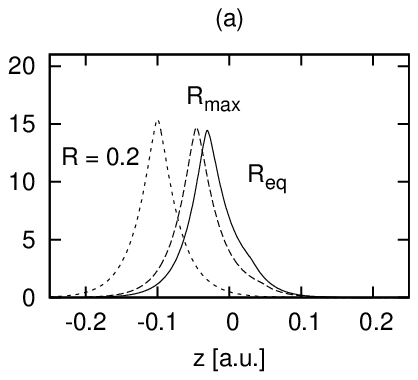}
\includegraphics[width=0.41\textwidth]{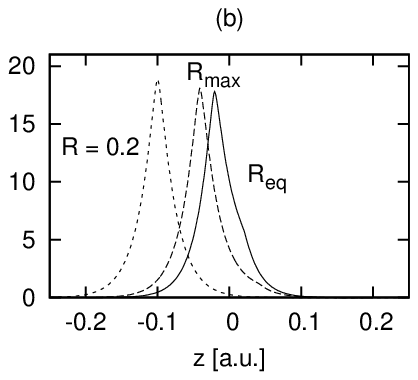}
\caption{$z$-profile of the electronic distribution $\int\int |\psi_{0}(x,y,z)|^{2}dx\,dy$ of the ground state of the molecular ion ${\rm LiH}^{3+}$ for internuclear distances: $R_{eq}$, $R_{max}$, $R=0.2$\ a.u.  and magnetic fields: (a) $B=10^{6}$\ a.u. and (b) $B=5\times 10^{6}$\ a.u.}
\label{ZDElih}
\end{figure}

A plot of the total energy $E_{t}$ as a function of the internuclear distance $R$ for $B=5\times10^{6}$\ a.u. is shown in figure~\ref{potlihB5e06}.
The ground state electronic distribution for a magnetic field $B=10^{6}$\ a.u. for three values of the internuclear distances: $R_{eq} \approx 0.06$, $R_{max} \approx 0.09$  and $R_{cross} \approx 0.12$\ a.u. is shown in figure~\ref{DE:lih}. At equilibrium, the electronic distribution displays  one peak which follows the position of the Li nucleus. Another view is presented in  figure~\ref{ZDElih} where  the integrated electronic distribution along the $z$-axis is displayed for three different values of the internuclear distance and for $B=10^{6}$\ a.u. and $B=5\times10^{6}$\ a.u. In the latter case,  the electronic distribution is almost symmetric around the position of the Li nucleus with a small asymmetry  due to the presence of the proton. When we compare both fields, we can see how the electronic distribution is shrunk along the $z$-axis when the magnetic field is increased.

\section{Conclusions}

We presented a non-relativistic study of the ground state of the one-electron Lithium-containing ions ${\rm Li}^{2+}$, ${\rm LiH}^{3+}$, ${\rm LiHe}^{4+}$, ${\rm Li}_{2}^{5+}$ in the presence of a strong magnetic field $B\le 10^{7}$~a.u in parallel configuration within the  Born-Oppenheimer approximation of zero order. Two methods were employed complementing each other: the {\it variational method}  and the {\it Lagrange-mesh method}.

The obtained results give clear indications that the exotic molecular ions
${\rm Li}_{2}^{5+}$, ${\rm LiHe}^{4+}$,  ${\rm LiH}^{3+}$ begin to exist as metastable states starting at the threshold magnetic fields $B_{th} \sim  2 \times 10^4, \sim 10^5$ and  $\sim 10^6$\ a.u., respectively.  As the magnetic field increases the potential wells of all three systems become sufficiently deep to keep more than one (longitudinal) vibrational state.
Eventually, the ions  ${\rm Li}_{2}^{5+}$, ${\rm LiHe}^{4+}$ become stable or almost stable at magnetic fields $B\simeq  10^{6}\,$ and $B\simeq 10^{7}$\ a.u., respectively, when the ion ${\rm LiH}^{3+}$ remains unstable towards decay ${\rm LiH}^{3+}   \rightarrow {\rm Li}^{2+} + p$ in the whole domain of magnetic fields considered
$B\leq10^{7}$\ a.u. The energies obtained with the variational method were improved with Lagrange-mesh calculations for magnetic fields $B\lesssim 10^6$~a.u. for all three molecular ions studied  (see tables  \ref{Tlili}, \ref{Tlihe} and \ref{Tlih}). For magnetic fields $B>10^6$~a.u., we were unable to reach converged results in the Lagrange-mesh calculations.

The accuracy of the variational energy can be tested using the converged Lagrange-mesh results. For $B\simeq  10^6$\ a.u., the relative accuracies on the binding energy are very close: $\sim 0.05\%$, $\sim 0.03\%$, $\sim 0.04\%$, for ${\rm Li}_{2}^{5+}$, ${\rm LiHe}^{4+}$,  ${\rm LiH}^{3+}$, respectively.

All these molecular ions have a similar behaviour: when the magnetic field increases, each molecular ion becomes more compact (the equilibrium distance decreases) and more bound (the binding energy increases). For a given magnetic field where all three molecular ions display a minimum in the corresponding potential curves that support at least one vibrational level, we find a hierarchy of the binding energies,
\begin{displaymath}
E_{b}^{{\rm Li}_{2}^{5+}} > E_{b}^{{\rm LiHe}^{4+}} > E_{b}^{{\rm LiH}^{3+}}\,,
\end{displaymath}
$i.e.$, the most bound lithium-containing  molecular system is ${\rm Li}_{2}^{5+}$.
If we consider other ions with two centers and one electron~\cite{chains2009}: ${\rm He}_{2}^{3+}$, ${\rm HeH}^{2+}$ and ${\rm H}_{2}^{+}$, in a magnetic field $10^6 \le~ B\le 10^7$\ a.u., there is a hierarchy of ionization energies
\begin{displaymath}
   E_{b}^{{\rm Li}_{2}^{5+}}\ >\ E_{b}^{{\rm LiHe}^{4+}} >\ E_{b}^{{\rm LiH}^{3+}}\ >\ E_{b}^{{\rm He}_{2}^{3+}}\ >\ E_{b}^{{\rm HeH}^{2+}}\ >\
   E_{b}^{{\rm H}_{2}^{+}} \ .
\end{displaymath}
The ionization energy grows with an increase of the total charge of the nuclei. In the case of the same charge, the higher binding energy corresponds to the presence of the nucleus with the higher charge.

\ack{HOP thanks the Universit\'e Libre de Bruxelles and the PNTPM Department for their hospitality during the final stage of the present work. HOP was supported in part by a CONACyT grant for PhD studies (M\'exico) and an FNRS grant (Belgium).}

\end{document}